\documentclass[10pt]{iopart}
\usepackage{iopams}
\usepackage{graphicx}    
\begin{document}

\title[Equilibrium conditions of spinning test particles in KdS spacetimes]{Equilibrium conditions of spinning test particles in Kerr-de~Sitter spacetimes}

\author{Zden\v{e}k Stuchl\'{i}k\footnote[1]{Zdenek.Stuchlik@fpf.slu.cz} and Ji\v{r}\'{i} Kov\'{a}\v{r}
\footnote[2]{Jiri.Kovar@fpf.slu.cz}}

\address{Institute of Physics, Faculty of Philosophy and Science, Silesian University in Opava, Bezru\v{c}ovo n\'{a}m. 13, 746 01 Opava, Czech Republic}

\begin{abstract}
Equilibrium conditions and spin dynamics of spinning test particles are discussed in the stationary and axially symmetric Kerr-de~Sitter black-hole or naked-singularity spacetimes. The general equilibrium conditions are established, but due to their great complexity, the detailed discussion of the equilibrium conditions and spin dynamics is presented only in the simple and most relevant cases of equilibrium positions in the equatorial plane and on the symmetry axis of the  spacetimes. It is shown that due to the combined effect of the rotation of the source and the cosmic repulsion the equilibrium is spin dependent in contrast to the spherically symmetric spacetimes. 
In the equatorial plane, it is possible at the so-called static radius, where the gravitational attraction is balanced by the cosmic repulsion, for the spinless particles as well as for spinning particles with arbitrarily large $\varphi$-oriented spin or at any radius outside the ergosphere with a specifically given spin orthogonal to the equatorial plane.
On the symmetry axis, the equilibrium is possible at any radius in the stationary region and is given by an appropriately tuned spin directed along the axis. At the static radii on the axis the spin of particles in equilibrium must vanish. 

\end{abstract}

\pacs{04.20.-q, 04.25.-q, 04.70.Bw}



\section{Introduction}
Motion of test particles describes in an illustrative way properties of black-hole and naked-singularity spacetimes. The motion of uncharged and spinless test particles is governed by  geodesic equations and directly determines the geodesic structure of the spacetimes. Charged test particles can test combined gravitational and electromagnetic field of these backgrounds, their motion is determined by the Lorentz equation. 
If the test particles possess also spin, their equations of motion are more complex in comparison with the spinless particles because of the interaction of spin with the curvature of the spacetime given by the Riemann tensor \cite{Pap:1951:PRRSLA:}, \cite{Pir:1956:ACTPA:}. Moreover, the spin dynamics has also to be considered. In the absence of an electromagnetic field of the background, the spin dynamics is determined by the Fermi-Walker transport equation.

Studies of the equilibrium positions (equilibrium hereinafter) of charged test particles give direct information on interplay of the gravitational and electromagnetic forces acting in the charged (Reissner-Nordstr\"{o}m and Kerr-Newman) backgrounds \cite{Bic-Stu-Bal:1989:BULAI:}, \cite{Bon:1993:CLAQG:}, \cite{Bal-Bic-Stu:1989:BULAI:}, \cite{Agu-etal:1995:CLAQG:}, \cite{Stu-Bic-Bal:1999:GENRG2:}. In the simplest Schwarzschild backgrounds, the equilibrium of test particles is impossible, because only the gravitational attraction is acting here. However, the presence of a repulsive cosmological constant allows the equilibrium of even uncharged particles. 
Further, it was shown that in the Schwarzschild-de Sitter (SdS) backgrounds, the equilibrium of spinning test particles is independent of the particle spin being restricted to the static radius, where the gravitational attraction is just balanced by the cosmic repulsion \cite{Stu:1999:ACTPS2:}. The equilibrium is spin-independent also in the Reissner-Nordstr\"{o}m-de~Sitter  spacetimes \cite{Stu-Hle:2001:PHYSR4:}.
Of course, the equilibrium is spin dependent in the rotating Kerr spacetimes due to the interaction of the spin of the particle and the black hole \cite{Agu-etal:1995:CLAQG:}.

We focus onto more complicated case of stationary and axially symmetric spacetimes around rotating black holes or naked singularities in the universe with the recently indicated repulsive cosmological constant, i.e., on Kerr-de~Sitter (KdS) spacetimes, in order to extend the preliminary studies and to better understand the combined effects of the rotation of the source and the cosmic repulsion. For comparison, the restriction of our results to the pure Kerr and SdS spacetimes is also included. Because of the complexity of general equilibrium conditions, the detailed discussion is restricted only to the cases of the equatorial plane and the symmetry axis of the spacetimes, similarly to the discussion presented in the study of equilibrium conditions in the Kerr-Newman spacetimes \cite{Agu-etal:1995:CLAQG:}. In \mbox{Section 2}, properties of the KdS spacetimes are summarized, and in \mbox{Section 3}, equations of motion of spinning particles and the spin dynamics are given. Equilibrium conditions are determined and discussed in \mbox{Section 4}, while concluding remarks are presented in \mbox{Section 5}. 
\section{Kerr-de~Sitter spacetimes}
KdS spacetimes are stationary and axially symmetric solutions of Einstein's equations with a non-zero cosmological constant $\Lambda$. 
\subsection{Geometry}
In the standard Boyer-Lindquist coordinates $(t,\varphi,r,\theta)$ and geometric units $(c=G=1)$, the line element of the KdS geometry is given by the relation
\begin{eqnarray}
\label{K1}
ds^{2}=\frac{a^2\Delta_\theta \sin^2{\theta}-\Delta_r}{I^2\rho^2}dt^2+2\,\frac{a\sin^2{\theta}[\Delta_r-(a^2+r^2)\Delta_\theta]}{I^2\rho^2}dtd\varphi+\\\nonumber
\frac{\sin^2{\theta}[(a^2+r^2)^2\Delta_\theta-a^2\Delta_r\sin^2{\theta}]}{I^2\rho^2}d\varphi^2+
\frac{\rho^2}{\Delta_r}dr^2+\frac{\rho^2}{\Delta_{\theta}}d\theta^2,
\end{eqnarray}
where
\begin{eqnarray}
\label{K2}
\Delta_r&=&r^2-2Mr+a^2
    -\frac{1}{3}{\Lambda}r^2(r^2+a^2),\\
\label{K3}
\Delta_{\theta}
  &=&1+\frac{1}{3}{\Lambda}a^2\cos^2{\theta},\\
\label{K4}
I&=&1+\frac{1}{3}{\Lambda}a^2,\\
\label{K5}
\rho^2&=&r^{2}+a^2\cos^{2}{\theta} 
\end{eqnarray}
and the mass $M$, specific angular momentum $a$, and cosmological constant $\Lambda$ are parameters of the spacetimes.  
For $\Lambda=0$, we obtain the line element of the axially symmetric Kerr geometry, while the case $a=0$ corresponds to the spherically symmetric SdS geometry.  
Using the dimensionless cosmological parameter $\lambda
=\frac{1}{3}\Lambda M^2$ and putting $M=1$, the coordinates $t$, $r$, the line element $ds$, and the parameter $a$ are expressed in units of $M$ and become dimensionless.   
However, the standard Boyer-Lindquist coordinates are not well behaved on the symmetry axis and can not be used for establishing of the equilibrium conditions in this special case. Therefore we also  use the Kerr-Schild coordinates introduced by the transformation 
\begin{eqnarray}
\label{36}
x&=&(r^2+a^2)^{1/2}\sin{\theta}\cos{\varphi}\\
y&=&(r^2+a^2)^{1/2}\sin{\theta}\sin{\varphi}\\ 
z&=&r\cos{\theta}.
\end{eqnarray}
Note that in the following discussion, some characteristic functions appear. As for the notation, the functions involving the variable $r$ are derived and related to the equatorial plane, while functions involving the variable $z$ are valid for the symmetry axis. 

\subsection{Black-hole and naked-singularity spacetimes}
The stationary regions of the KdS spacetimes, determined by the relation $\Delta_r(r;a^2,\lambda)\geq 0$, are limited by the inner and outer black-hole horizons at $r_{h-}$ and $r_{h+}$ and by the cosmological horizon at $r_{c}$. Spacetimes containing three horizons are black-hole (BH) spacetimes, while spacetimes containing one horizon (the cosmological horizon exists for any choice of the spacetime parameters) are naked-singularity (NS) spacetimes \cite{Stu-Sla:2004:PHYSR4:}. 

\begin{figure}
\begin{center}
\includegraphics[width=1.05\hsize]{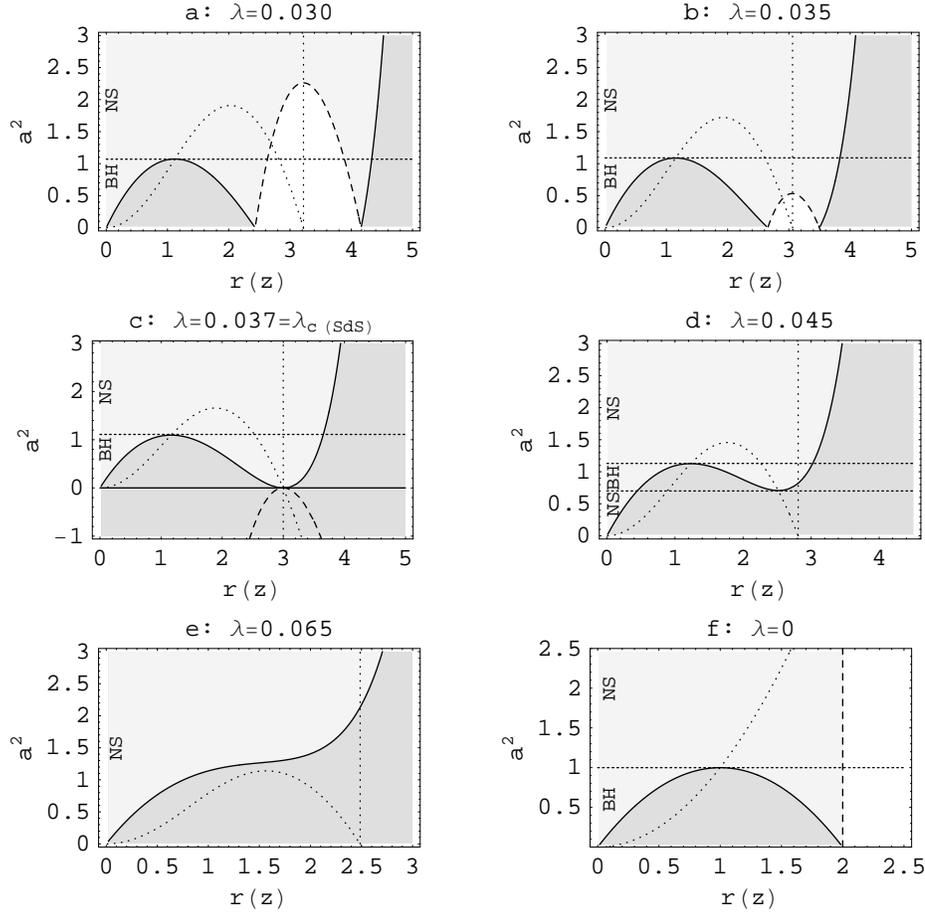}
\end{center}
\caption{Location of event horizons, static limit surfaces, and static radii in the KdS spacetimes. The coalescing functions $a^2_{h}(r;\lambda)$ and $a^2_{h}(z;\lambda)$ (solid) determine loci of horizons in the equatorial plane as well as on the symmetry axis. They also separate the dynamic regions (dark gray) and the stationary regions (light gray and white) of the spacetimes. The function $a^2_{sl}(r;\lambda)$ (dashed) determines the radii of static limit surfaces in the equatorial plane and separates the ergosphere (light gray) and the other stationary regions in the equatorial plane of the spacetimes (white). The function $a^2_{sr}(z;\lambda)$ (dotted) determines loci of static radii on the symmetry axis. The vertical dotted lines denote positions of the static radii in the equatorial plane and the vertical dashed line denotes the position of the static limit surface in the equatorial plane of the Kerr spacetimes. We give examples of different types of behaviour of the functions in dependence on the value of $\lambda$ discussed in the text. Figures (a)-(e) concern the KdS spacetimes, and for comparison, the figure (f) concerns the Kerr spacetimes. Note that in the figure (c), the irrelevant negative function $a^2_{sl}(r;\lambda)$ is exceptionally illustrated.}
\label{Fig:1}
\end{figure}

For given values of the spacetime parameters $a^2$ and $\lambda$, the radii of the horizons are, due to the relation $\Delta_r(r;a^2,\lambda)=0$, determined by solutions of the equation 
\begin{eqnarray}
\label{K6}
a^2=a^2_h(r;\lambda)\equiv\frac{r^2-2r-\lambda r^4}{\lambda r^2-1}.
\end{eqnarray}
For a fixed value of $\lambda$, the number of solutions of this equation  (horizons) depends  on the number of positive local extrema of the function $a^2_h(r;\lambda)$ (see Figure~\ref{Fig:1}). The extrema are located at the radii implicitly determined (due to the condition $\partial_r(a^2_h(r;\lambda)=0$) by the relation 
\begin{eqnarray}
\label{K6a}
\lambda=\lambda_{he}(r)\equiv\frac{2r+1-\sqrt{8r+1}}{2r^3},
\end{eqnarray}
whereas the maximum of the function $\lambda_{he}(r)$ is located at $r_{c}=(3+2\sqrt{3})/4$ and  takes the critical value $\lambda_{c(KdS)}\doteq 0.05924$.   
The behaviour of the function $a^2_h(r;\lambda)$ can be summarized in the following way 
\begin{itemize}    
\item[$\bullet$] $0<\lambda<\lambda_{c(KdS)}$, there are two local extrema of the function $a^2_h(r;\lambda)$, denoted as $a^2_{h,max}(\lambda)$ and $a^2_{h,min}(\lambda)$ and determined by the relations (\ref{K6}) and (\ref{K6a}), whereas the local minimum $a^2_{h,min}(\lambda)$ becomes positive (relevant) for $\lambda>\lambda_{c(SdS)}\equiv1/27\doteq0.03704$. Thus for $a^2>a^2_{h,max}(\lambda)$ or for $a^2<a^2_{h,min} (\lambda)$, there is one solution of the equation (\ref{K6}) and only NS spacetimes exist. For $a^2<a^2_{h,max}(\lambda)$ and $a^2>a^2_{h,min}(\lambda)$, there are three solutions of the equation (\ref{K6}) and BH spacetimes exist (see \mbox{Figures~\ref{Fig:1}a-d)}.
\item[$\bullet$] $\lambda=\lambda_{c(KdS)}$, the local extrema $a^2_{h,max}(\lambda)$ and $a^2_{h,min}(\lambda)$ coalesce at $r_{c}$ and take the value $a^2_{c}\doteq 1.21202$. Then there is only one solution of the equation (\ref{K6}) for all $a^2>0$ and only NS spacetimes exist.
\item[$\bullet$] $\lambda>\lambda_{c(KdS)}$, there are no extrema of the function $a^2_h(r;\lambda)$ and thus there is only one solution of the equation (\ref{K6}) for all $a^2>0$ and only NS spacetimes exist  (see Figure~\ref{Fig:1}e).
\end{itemize}
The degenerate cases corresponding to extreme black holes or naked singularities are discussed in \cite{Stu-Sla:2004:PHYSR4:}. 

In the case of the Kerr spacetimes, there exist BH spacetimes for $a^2\leq1$ and NS spacetimes for $a^2>1$ (see Figure~\ref{Fig:1}f) \cite{Mis-Tho-Whe:1973:Gra:} and in the case of the SdS spacetimes, there exist only BH spacetimes for $\lambda\leq\lambda_{c(SdS)}$, whereas for $\lambda>\lambda_{c(SdS)}$, the SdS spacetimes are dynamic \cite{Stu-Hle:1999:PHYSR4:}. 

The discussion of the position of the horizons is $\theta$-independent, i.e., it is valid also for the symmetry axis. The horizons on the symmetry axis are then determined by the equation 
\begin{eqnarray}
\label{K6b}
a^2=a^2_h(z;\lambda)\equiv\frac{z^2-2z-\lambda z^4}{\lambda z^2-1},
\end{eqnarray}
arriving from the relation (\ref{K6a}) simply after replacing $r$ by $z$. Restricting here and hereafter to the positive values of $z$, we can interpret them as "radii" on the symmetry axis  because of the axial symmetry of the spacetimes.  

\subsection{Static radii}
Due to the repulsive cosmological constant and the rotation of the source, the KdS spacetimes enable existence of the so-called static radii, where the "free" spinless particles can "stay at rest". A detailed discussion of the properties and astrophysical relevance of the static radii can be found in \cite{Stu-Sla:2004:PHYSR4:}. 
Apparently, these radii are defined by the \mbox{4-velocity} (\ref{11}) and the equation of motion (\ref{7}), reduced to the geodesic equation in this case. 

In the equatorial plane, the static radius is located at 
\begin{eqnarray}
\label{K7a}
r=r_{sr}\equiv \lambda^{-1/3},
\end{eqnarray}
where the gravitational attraction is balanced by the cosmological repulsion independently of the rotational parameter $a$. It plays important role in the geodetical structure of the spacetimes \cite{Stu-Sla:2004:PHYSR4:} and even relates to the equilibrium of spinning test particles.

On the symmetry axis, there are at most two static radii dependent on both spacetime parameters $a$ and $\lambda$. For given values of the parameters, the static radii are implicitly determined by solutions of the equation 
\begin{eqnarray}
\label{45}
a^2=a^2_{sr}(z;\lambda)\equiv\frac{-1-2z^3\lambda+\sqrt{1+8z^3\lambda}}{2z\lambda}.
\end{eqnarray}
There is always one maximum of the function located at   
\begin{eqnarray}
z=2^{-2/3}\lambda^{-1/3}
\end{eqnarray}
taking the value
\begin{eqnarray}
\label{45a}
a^2_{sr,max}(\lambda)=\frac{2\sqrt{3}-3}{2^{4/3}\lambda^{2/3}}.
\end{eqnarray}
Note that the static radii on the axis are relevant for our discussion only in the stationary regions of the spacetimes as well as in the case of the equatorial plane. Then finding the common points of the functions $a^2_{sr}(z;\lambda)$ and $a^2_h(z;\lambda)$ at the extrema of the function $a^2_h(z;\lambda)$ and  summarizing other behaviour of the functions, we can conclude that in the case of 
\begin{itemize}    
\item[$\bullet$] $0<\lambda<\lambda_{c(KdS)}$, there is $a^2_{sr,max}(\lambda)>a^2_{h,max}(\lambda)$, therefore, on the symmetry axis of the NS spacetimes with $a^2>a^2_{h,max}(\lambda)$, there is no static radius for $a^2>a^2_{sr,max}(\lambda)$, one static radius for $a^2=a^2_{sr,max}(\lambda)$, and two static radii for $a^2_{h,max}(\lambda)<a^2<a^2_{sr,max}(\lambda)$. In the BH spacetimes with $a^2_{h,min}(\lambda)<a^2<a^2_{h,max}(\lambda)$, there is only one relevant static radius and in the NS spacetimes with $a^2<a^2_{h,min}(\lambda)$, there are no relevant static radii on the axis (see Figures~\ref{Fig:1}a-d).
\item[$\bullet$] $\lambda\geq\lambda_{c(KdS)}$, there is $a^2_{sr}(z;\lambda)\leq a^2_{h}(z;\lambda)$, thus there are no relevant static radii on the symmetry axis for any KdS spacetimes (see Figure~\ref{Fig:1}e).
\end{itemize}

In the case of the Kerr spacetimes, there are no static radii in the BH spacetimes neither in the equatorial plane nor on the symmetry axis, but in the NS spacetimes, one static radius appears on the axis (see Figure~\ref{Fig:1}f). In the case of the SdS spacetimes, there is one static radius at $r_{sr}=\lambda^{-1/3}$, of course, due to the spherical symmetry, for any value of $\theta$.

\subsection{Static limit surfaces and ergosphere}
For our purposes, i.e., determination of equilibrium positions of particles, it is necessary to determine behaviour of the ergosphere (for definition see, e.g., \cite{Mis-Tho-Whe:1973:Gra:}). The ergosphere of the KdS spacetimes, determined by the relation $g_{tt}>0$, is limited by the static limit surfaces, given by the equation $g_{tt}=0$.
In the equatorial plane, the radii of these surfaces $r_{sl-}$ (inner) and $r_{sl+}$ (outer) are given by solutions of the equation
\begin{eqnarray}
\label{K7}
a^2=a^2_{sl}(r;\lambda)\equiv\frac{r-2-\lambda r^3}{\lambda r}. 
\end{eqnarray}
For fixed value of $\lambda$, the possible number of solutions of this equation (static limit surfaces) depends on the number of positive local extrema of the function $a^2_{sl}(r;\lambda)$ (see Figure~\ref{Fig:1}). We can find (due to the condition $\partial_r(a^2_{sl}(r;\lambda)=0$) only one extremum of the function located at the static radius and taking the value
\begin{eqnarray}
a^2_{sl,max}(\lambda)=\lambda^{-2/3}(\lambda^{-1/3}-3).
\end{eqnarray}
There is $a^2_{sl,max}(\lambda_{c(SdS)})=0$, while it becomes positive for $\lambda<\lambda_{c(SdS)}$ and diverges for $\lambda\rightarrow0$. 
Further, we have to distinguish the cases when the static limit surfaces do exist or do not exist in the black-hole spacetimes. The critical values of the rotational and cosmological parameters are given by the relation 
\begin{eqnarray}
\label{K7c}
a^2_{h,max}(\lambda)=a^2_{sl,max}(\lambda)
\end{eqnarray}
and found to be $a^2_{e,BH}\doteq1.08317$ and $\lambda_{e,BH}\doteq0.03319$. 
Considering the asymptotic behaviour and the zero points of the function $a^2_{sl}(r;\lambda)$, we can summarize the number of static limit surfaces in the equatorial plane of the KdS spacetimes. In the case of 
\begin{itemize}
\item[$\bullet$] $\lambda<\lambda_{c(SdS)}$, there are two static limit surfaces for $a^2<a^2_{sl,max}(\lambda)$, one static limit surface for $a^2=a^2_{sl,max}(\lambda)$, and none static limit surface for $a^2>a^2_{sl,max}(\lambda)$ (see Figures~\ref{Fig:1}a,b). 
If $\lambda<\lambda_{e,BH}$, the static limit surface exists in all the black-hole spacetimes, and in naked-singularity spacetimes with $a^2\leq a^2_{sl,max}(\lambda)$ and it does not exist for $a^2>a^2_{sl,max}(\lambda)$ in the naked-singularity spacetimes (see Figure~\ref{Fig:1}a).
If $\lambda_{e,BH}<\lambda<\lambda_{c(SdS)}$, the static limit surface exists for black-hole spacetimes with $a^2\leq a^2_{sl,max}(\lambda)$ and do not exist for the black-hole and naked-singularity spacetimes with $a^2>a^2_{sl,max}(\lambda)$ (see Figure~\ref{Fig:1}b). 
\item[$\bullet$] $\lambda\geq\lambda_{c(SdS)}$, there are no static limit surfaces for all $a^2>0$ (see Figures~\ref{Fig:1}c-e). 
\end{itemize}

On the symmetry axis, the static limit surfaces coalesce with the horizons, i.e., they are determined by the function $a^2_h(z;\lambda)$ (\ref{K6}) and thus there is no ergosphere on the axis. 

In the equatorial plane of the Kerr spacetimes, there is the only static limit surface at $r=2$, independently of $a^2$, whereas on the symmetry axis, the surface coalesce with the outer event horizon (see Figure~\ref{Fig:1}f). In the case of the SdS spacetimes, there is no static limit surface.

\subsection{Classification}
\begin{figure}
\begin{center}\includegraphics[width=0.7\hsize]{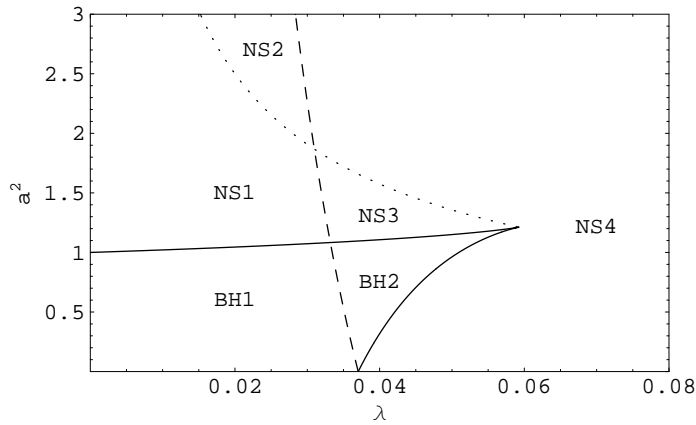}\end{center}
\caption{Classification of the KdS spacetimes. The parametric plane $(\lambda,a^2)$ is divided by the functions $a^2_{h,max}(\lambda)$ (upper solid), $a^2_{h,min}(\lambda)$ (lower solid), $a^2_{sl,max}(\lambda)$ (dashed), and $a^2_{sr,max}(\lambda)$ (dotted) into six regions corresponding to the classes of the KdS spacetimes BH1-NS4 differing in the number of horizons, static limit surfaces, static radii in the equatorial plane and on the symmetry axis.}     
\label{Fig:2}
\end{figure}

According to the given discussion presented above, the KdS spacetimes can be divided into six classes (see Table \ref{Tab:1} and Figure~\ref{Fig:2}) differing in the number of horizons, static limit surfaces, static radii in the equatorial plane and on the symmetry axis. 
\begin{table}
\caption{\label{Tab:1} Classification of KdS spacetimes according to the number of horizons, static limit surfaces, static radii in the equatorial plane and on the symmetry axis. In each column, the first of two digits denotes the number in the equatorial plane, the second one on the symmetry axis.}
\begin{indented}
\item\begin{tabular}{@{}cccc}
\br
\bf{Class}&\bf{event}&\bf{static limit}&\bf{static}\\
&\bf{horizons}&\bf{surfaces}&\bf{radii}\\
\hline
BH1&3,3&2,3&1,1\\
BH2&3,3&0,3&0,1\\
NS1&1,1&2,1&1,2\\
NS2&1,1&2,1&1,0\\
NS3&1,1&0,1&0,2\\
NS4&1,1&0,1&0,0\\
\br
\end{tabular}
\end{indented}
\end{table}
\section{Equations of motion and spin dynamics}
The motion of a spinning test particle of mass $m$ with a \mbox{4-velocity} $u^{\lambda}$ and spin tensor $S^{\mu\nu}$ in an arbitrary gravitational field has been studied by Papapetrou \cite{Pap:1951:PRRSLA:}. Such particle deviates from its geodesic motion and moves along a different orbit due to the spin-curvature interaction. Introducing the Pirani spin supplementary condition \cite{Pir:1956:ACTPA:} 
\begin{eqnarray}
\label{0}
S^{\mu\nu}u_\nu=0
\end{eqnarray}
and the covariant spin vector 
\begin{eqnarray}
\label{0a}
S_\sigma=\frac{1}{2}\epsilon_{\rho\mu\nu\sigma}u^{\rho}S^{\mu\nu},
\end{eqnarray}
the motion is governed by the equation
\begin{eqnarray}
\label{7}
m\frac{Du^{\alpha}}{d\tau}&=&-\epsilon^{\alpha\mu\nu\beta}\frac{D^2u_{\beta}}{d\tau^2}S_{\mu}u_{\nu}+\frac{1}{2}\epsilon^{\lambda\mu\rho\sigma}R^{\alpha}_{\nu\lambda\mu}u^{\nu}u_{\sigma}S_{\rho},  
\end{eqnarray}
where $\epsilon_{\rho\mu\nu\sigma}$ is the Levi-Civita completely antisymmetric tensor, $D/d\tau$ denotes the covariant derivate along the vector field $u^\alpha$, i.e.,
\begin{eqnarray}
\label{8}
\frac{Du^{\alpha}}{d\tau}&=&u^\beta(\partial_\beta u^\alpha+\Gamma^\alpha_{\beta \gamma}u^\gamma),
\end{eqnarray}
whereas $\Gamma^\alpha_{\beta\gamma}$ denotes coefficients of the affine connection of the background and $R^{\alpha}_{\nu\lambda\mu}$ is the Riemann tensor describing the background on which particle moves. The test particle is assumed to be so small in size and in mass not to modify the background. Clearly, we obtain the geodesic motion in the case of spinless particles. 
By construction of the spin vector (\ref{0a}), $S^\sigma$ is permanently orthogonal to the \mbox{4-velocity} $u^\sigma$, i.e.,
\begin{eqnarray}
\label{9}
S^\sigma u_\sigma&=&0.
\end{eqnarray}
Dynamics of the spin vector is then given by a relatively simple equation of the Fermi-Walker transport  
\begin{eqnarray}
\label{10}
\frac{DS_\alpha}{d\tau}&=&u_\alpha \frac{Du^\beta}{d\tau}S_\beta.
\end{eqnarray}

In order to consider the equilibrium of a spinning test particle, we must find conditions which guarantee that the equations of motion (\ref{7}) and the spin dynamics equations (\ref{10}), along with the orthonomality relation (\ref{9}), are simultaneously satisfied for the \mbox{4-velocity} $u^\alpha$ corresponding to a stationary particle having the only non-zero time component, given by the relations
\begin{eqnarray}
\label{11}
u^\alpha=\frac{1}{\sqrt{-g_{tt}}}\delta^t_\alpha,\quad \frac{du^\alpha}{d\tau}=u^\beta\partial_\beta u^\alpha=0, 
\end{eqnarray}
in the background under consideration.
Since the first derivate of this \mbox{4-velocity} reduces to the form
\begin{eqnarray}
\label{27}
\frac{Du^{\alpha}}{d\tau}&=&(u^t)^2\Gamma^\alpha_{tt}
\end{eqnarray}
and the orthogonality of the spin and the \mbox{4-velocity} implies that 
\begin{eqnarray}
S_t=0, 
\end{eqnarray}
i.e., only space components of the spin vector are non-zero, the spin dynamics equation (\ref{10}) reduces to the form
\begin{eqnarray}
\label{15}
\frac{DS_\alpha}{d\tau}&=&[u_\alpha \Gamma^i_{tt}(u^t)^2]S_i,
\end{eqnarray}
which implies
\begin{eqnarray}
\label{16}
\frac{dS_\alpha}{d\tau}&=&[\Gamma^i_{\alpha t}u^t+u_\alpha \Gamma^i_{tt}(u^t)^2]S_i.
\end{eqnarray}
Moreover, notice that the second derivate of the \mbox{4-velocity} can be now rewritten as
\begin{eqnarray}
\label{22}
\frac{D^2u_\alpha}{d\tau^2}=(u^t)^2\Gamma^\beta_{t\alpha}\Gamma^\gamma_{\beta t}u_\gamma.
\end{eqnarray}
\section{Equilibrium conditions}   
The basic equilibrium condition for spinning test particles follows immediately from the \mbox{4-velocity} of such particles given by the equation (\ref{11}). The equilibrium of spinning test particles is possible only outside ergosphere, where $g_{tt}<0$, whereas such regions occur only in the stationary regions, where $\Delta_r>0$. Together with the condition $S_t=0$, it is the   general limiting condition. 
Other equilibrium conditions following from the spin dynamics equations (\ref{16}) and the equation of motion (\ref{7}), assuming the \mbox{4-velocity} derivatives in the form (\ref{27}) and (\ref{22}), are discussed in details separately in the case of the equatorial plane and the symmetry axis.  

\subsection{General equilibrium conditions}
Outside the symmetry axis, the spin dynamics equations (\ref{16}) can be written in the general form 
\begin{eqnarray}
\label{17}
\frac{dS_t}{d\tau}&=&0,\quad \frac{dS_r}{d\tau}=u^t\Gamma^\varphi_{rt}S_\varphi,\quad \frac{dS_\theta}{d\tau}=u^t\Gamma^\varphi_{\theta t}S_\varphi,\\
\label{20}
\frac{dS_\varphi}{d\tau}&=&u^t[(\Gamma^r_{\varphi t}-\frac{g_{t\varphi}}{g_{tt}}\Gamma^r_{tt})S_r+(\Gamma^\theta_{\varphi t}-\frac{g_{t\varphi}}{g_{tt}}\Gamma^\theta_{tt})S_\theta].
\end{eqnarray}
Since the \mbox{4-velocity} derivatives given in the equations (\ref{27}) and (\ref{22}) take the  form
\begin{eqnarray}
\label{28}
\frac{Du^{t}}{d\tau}=0,\quad
\frac{Du^{\varphi}}{d\tau}=0,\quad
\frac{Du^{r}}{d\tau}=(u^t)^2\Gamma^r_{tt},\quad
\frac{Du^{\theta}}{d\tau}=(u^t)^2\Gamma^\theta_{tt},
\end{eqnarray}
\begin{eqnarray}
\label{23}
\frac{D^2u_t}{d\tau^2}&=&-u^t[(\Gamma^r_{tt}\Gamma^t_{rt}+\Gamma^\theta_{tt}\Gamma^t_{\theta t})+\frac{g_{t\varphi}}{g_{tt}}(\Gamma^r_{tt}\Gamma^\varphi_{rt}+\Gamma^\theta_{tt}\Gamma^\varphi_{\theta t})],\\
\label{24}
\frac{D^2u_\varphi}{d\tau^2}&=&-u^t[(\Gamma^r_{\varphi t}\Gamma^t_{rt}+\Gamma^\theta_{\varphi t}\Gamma^t_{\theta t})+\frac{g_{t\varphi}}{g_{tt}}(\Gamma^r_{\varphi t}\Gamma^\varphi_{rt}+\Gamma^\theta_{\varphi t}\Gamma^\varphi_{\theta t})],\\
\label{25}
\frac{D^2u_r}{d\tau^2}&=&0,\quad \frac{D^2u_\theta}{d\tau^2}=0,
\end{eqnarray}
the equation of motion (\ref{7}) implies four equilibrium conditions in the implicit form
\begin{eqnarray}
\label{28a}
R^t_{t\theta r}S_\varphi u^t u_t=0,\\
\label{28aa}
R^\varphi_{t\theta r} S_\varphi u^t u_t=0,\\
\label{28b}
S_r(u_t R^r_{t\theta\varphi}-u_{\varphi}R^r_{t\theta t})-m u^t\sqrt{-g}\Gamma^r_{tt}-S_{\theta}[u_t R^r_{tr\varphi}+u_{\varphi}R^r_{trt}+A]=0,\\
\label{28c}
S_{\theta}(u_t R^\theta_{t\varphi r}-u_{\varphi}R^\theta_{ttr})-m u^t\sqrt{-g}\Gamma^\theta_{tt}-S_r[u_t R^{\theta}_{t\varphi\theta}+u_{\varphi}R^{\theta}_{t\theta t}+A]=0,
\end{eqnarray}
where 
\begin{eqnarray}
A=u^t [u_t^2(\Gamma^t_{rt}\Gamma^r_{t\varphi}+\Gamma^t_{\theta t}\Gamma^\theta_{t\varphi})-u_\varphi^2(\Gamma^\varphi_{rt}\Gamma^r_{tt}+\Gamma^\varphi_{\theta t}\Gamma^\theta_{tt})-\\\nonumber
u_t u_\varphi(\Gamma^t_{rt}\Gamma^r_{tt}-\Gamma^\varphi_{rt}\Gamma^r_{t\varphi}+\Gamma^t_{\theta t}\Gamma^\theta_{tt}-\Gamma^\varphi_{\theta t}\Gamma^\theta_{t\varphi})]
\end{eqnarray}
and $g$ is the determinant of the metric tensor. 
The coefficients of the affine connection and the components of the Riemann tensor of the KdS spacetimes can be readily computed with any computer algebra system. 

The conditions (\ref{28a})-(\ref{28c}) and the spin dynamics equations (\ref{17}) and (\ref{20}) are too long to be explicitly written and discussed in the general case. Therefore we restrict our attention to the most important cases of the equatorial plane and the symmetry axis, which give, moreover, relatively simple results. 
\subsection{Equilibrium in equatorial plane}
\begin{figure}
\begin{center}\includegraphics[width=1.05\hsize]{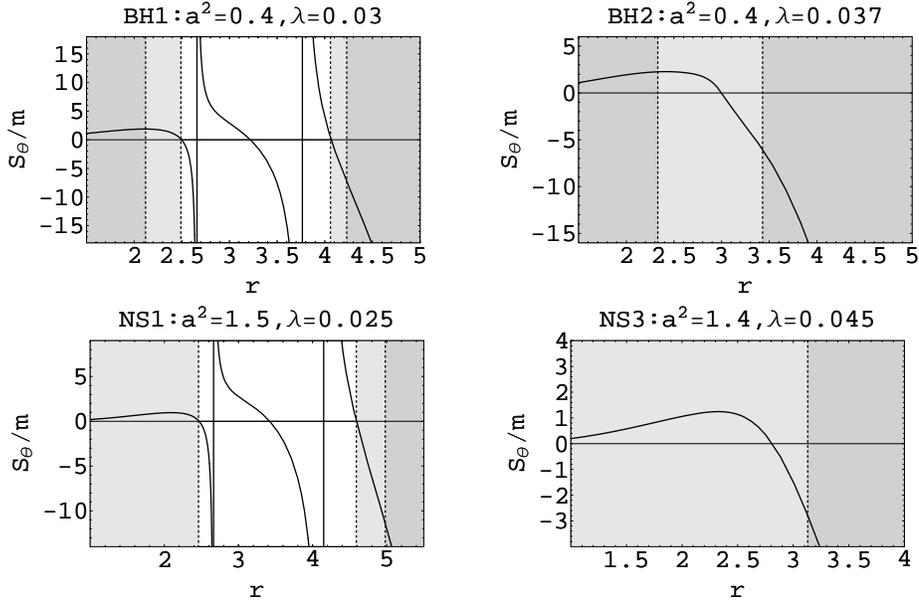}\end{center}
\caption{Behaviour of the function $S_{\theta}(r;a,\lambda)$. The function determines the magnitude and orientation of the latitudinal component of the spin vector of spinning particle in equilibrium at the radius $r$ for given values of the parameters $a^2$ and $\lambda$ in the case of $S_{\varphi}=0$. It vanishes at the radii of static limit surfaces, denoted by the dashed lines between the light gray and white regions, and at the static radius. The horizons of the spacetimes are denoted by the dashed lines between the dark gray and light gray regions. The divergence of the function is denoted by the solid vertical lines. We give four examples of the behaviour of the function for values of the parameters corresponding to the given classification of the KdS spacetimes (see Figure~\ref{Fig:2}). Note that the function $S_{\theta}(r;a,\lambda)$ is not relevant in the dynamic regions (gray).}
\label{Fig:3}
\end{figure}

As we noticed, the equatorial equilibrium is possible in the spacetimes admitting static limit surfaces, and outside the ergosphere, i.e., in the region where 
\begin{eqnarray}
\label{28d}
a^2<a^2_{sl}(r;\lambda), 
\end{eqnarray}
equivalently $a^2-\Delta_r<0$ or $r_{sl-}<r<r_{sl+}$.
Therefore we assume $\lambda<\lambda_{crit(SdS)}$ and $a^2<a^2_{sl,max}(\lambda)$.  
Now the spin dynamics equations (\ref{17}) and (\ref{20}) take the form
\begin{eqnarray}
\label{29}
\frac{dS_t}{d\tau}&=&\frac{dS_\theta}{d\tau}=0,\\
\label{30}
\frac{dS_r}{d\tau}&=&u^t\frac{a(1-r^3\lambda)}{r^2\Delta_r}S_\varphi,\\
\label{30a}
\frac{dS_\varphi}{d\tau}&=&u^t\frac{a(1-r^3\lambda)\Delta_r}{r^2I^2(a^2-\Delta_r)}S_r,
\end{eqnarray}
the equations (\ref{28a}), (\ref{28aa}) vanish, and the conditions (\ref{28b}), (\ref{28c}) yield      
\begin{eqnarray}
\label{32}
S_\theta=m\frac{r^2(1-\lambda r^3)(a^2-\Delta_r)}{a[\lambda^2r^6+\lambda(r^3+3a^2r)-3r+7](1+\lambda a^2)},
\end{eqnarray}
\begin{eqnarray}
\label{33}
\frac{S_ra\Delta_r[\lambda^2r^6-\lambda(5r^3+3ra^2)+3r-5]}{r^4(a^2-\Delta_r)^2}=0.
\end{eqnarray}
Therefore, outside the ergosphere, in the equatorial plane of the KdS spacetimes, the equilibrium of spinning test particles requires, due to the equation (\ref{29}), $S_\theta=const$ given by the equation (\ref{32}).  The conditions (\ref{30}), (\ref{30a}), and (\ref{33}) can be discussed in the following way. In the case of 
\begin{itemize}
\item[$\bullet$] $S_r=0$, the condition (\ref{33}) is automatically satisfied. The equation (\ref{30a}) requires  $S_{\varphi}=const$ and the condition (\ref{30}), where $dS_r /d\tau=0$, implies that for
\begin{itemize}
\item[$\bullet$] $S_{\varphi}\neq0$, the equilibrium is possible only at the static radius $r_{sr}$ (only if $r_{sr}$ satisfies the condition (\ref{28d})) with the spin $S_{\theta}$ given by the equation (\ref{32}), i.e., $S_\theta=0$
\item[$\bullet$] $S_{\varphi}=0$, the equilibrium is possible at all the radii satisfying the conditions (\ref{28d}) with the spin given by the function $S_{\theta}(r;a,\lambda)$ determined by the equation (\ref{32}). In the case of spinless particles ($S_\theta=0$), the equilibrium is possible at the static radius only and there is no equilibrium possible at the radii, where the function $S_{\theta}$ diverges. Note that in the case of $a^2=a^2_{sl,max}(\lambda)=\lambda^{-2/3}(\lambda^{-1/3}-3)$, the condition (\ref{32}) would allow the equilibrium independent of the spin $S_\theta$ at $r_{sr}$. But this is the limit case of the spacetimes which does not satisfy the condition $a^2<a^2_{sl}(r;\lambda)$. The static radius $r_{sr}$ is the radius where the static limit surfaces coalesce for $a^2=a^2_{sl,max}(\lambda)$ (see Figures \ref{Fig:1}a,b).    
We give the behaviour of the function $S_{\theta}(r;a,\lambda)$ for a few specifically chosen values of the spacetime parameters in Figure~\ref{Fig:3}. 
\end{itemize} 
\item[$\bullet$] $S_r\neq0$, the condition (\ref{33}) is satisfied at the radii given by solutions of the equation          
\begin{eqnarray}
\label{35}
a^2=a^2_{eq}(r;\lambda)\equiv \frac{\lambda^2r^6-5\lambda r^3+3r-5}{3\lambda r}.
\end{eqnarray}
But there is $a^2_{eq}(r;\lambda)\geq a^2_{sl}(r;\lambda)$ for all positive values of $r$, thus there is no equilibrium in this case.
\end{itemize}
\subsubsection{Kerr and Schwarzschild-de Sitter cases}
In the equatorial plane of the Kerr spacetimes, the equilibrium of spinning test particles is again possible only outside the ergosphere, i.e., in the region with $r>2$. The limit case ($\lambda=0$) of the equilibrium conditions (\ref{29})-(\ref{33}) implies that the equilibrium requires $S_\theta=const$, as well as in the KdS spacetimes. In the case of
\begin{itemize}  
\item[$\bullet$]$S_r=0$, the condition (\ref{33}) is automatically satisfied. The condition (\ref{30a}) implies $S_\varphi=const$ and the condition (\ref{30}), where $dS_r /d\tau=0$, is satisfied only in the case of $S_{\varphi}=0$. Then the equilibrium is possible at all the radii $r>2$ with spin given by the relation 
\begin{eqnarray}\label{35a}S_{\theta}=m\frac{r^3(2-r)}{a(7-3r)}.\end{eqnarray}
Note that there is no equilibrium possible for $S_{\theta}=0$ (spinless case), because of the restriction $r>2$.   
\item[$\bullet$] $S_r\neq0$, the condition (\ref{33}) implies the solution $r=5/3$, which  does not satisfy the condition $r>2$ and then there is no equilibrium in this case.
\end{itemize}
In the SdS spacetimes, the limit case ($a=0$) of the equilibrium conditions (\ref{29})-(\ref{33}) implies that the equilibrium is possible only at the static radius. The spin can be arbitrary and it will be time independent. Of course, because of the the spherical symmetry of the SdS spacetimes, this result holds for any central plane of the spacetime.

\subsection{Equilibrium on axis of symmetry}
On the symmetry axis, the equilibrium is possible in the regions where
\begin{eqnarray}
\label{36a}
a^2>a^2_h(z;\lambda).
\end{eqnarray}  
In the Kerr-Schild coordinates (\ref{36}), the spin dynamics equations (\ref{16}) can be written in the form 
\begin{eqnarray}
\label{39}
\frac{dS_t}{d\tau}&=&\frac{dS_z}{d\tau}=0,\\
\label{40}
\frac{dS_x}{d\tau}&=&-\omega S_y,\\
\label{40a}
\frac{dS_y}{d\tau}&=&\omega S_x,
\end{eqnarray}
where
\begin{eqnarray}
\omega=\frac{a[2z+(a^2+z^2)^2\lambda](1+a^2\lambda)\sqrt{a^2+z^2}}{(a^2+z^2)^2(1+a^2\lambda)\sqrt{\Delta_z}}
\end{eqnarray}
and using the \mbox{4-velocity} derivatives (\ref{27}) and (\ref{22}) in the form
\begin{eqnarray}
\frac{Du^t}{d\tau}=0,\quad \frac{Du^x}{d\tau}=0, \quad \frac{Du^y}{d\tau}=0, \quad
\frac{Du^z}{d\tau}=(u^t)^2\Gamma^z_{tt},
\label{38}
\end{eqnarray}
\begin{eqnarray}
\label{37}
\frac{D^2u_t}{d\tau^2}=-u^t\Gamma^z_{tt}\Gamma^t_{zt},\quad
\frac{D^2u_x}{d\tau^2}=0,\quad \frac{D^2u_y}{d\tau^2}=0, \quad \frac{D^2u_z}{d\tau^2}=0,
\end{eqnarray}
we obtain from the equation of motion (\ref{7}) three relations 
\begin{eqnarray}
\label{42}
\frac{a(a^2-3z^2)}{(a^2+z^2)^3(1+a^2\lambda)}S_x&=&0,\\
\label{42a}
\frac{a(a^2-3z^2)}{(a^2+z^2)^3(1+a^2\lambda)}S_y&=&0,
\end{eqnarray}
\begin{eqnarray}
\label{44}
S_z=-m\frac{(a^2+z^2)^2(1+a^2\lambda)^2(a^2-z^2+z(a^2+z^2)^2\lambda)}{2a\Delta_z(a^2-3z^2)},
\end{eqnarray}
whereas the equation (\ref{7}) for $\alpha=t$ vanishes.

\begin{figure}
\begin{center}\includegraphics[width=1\hsize]{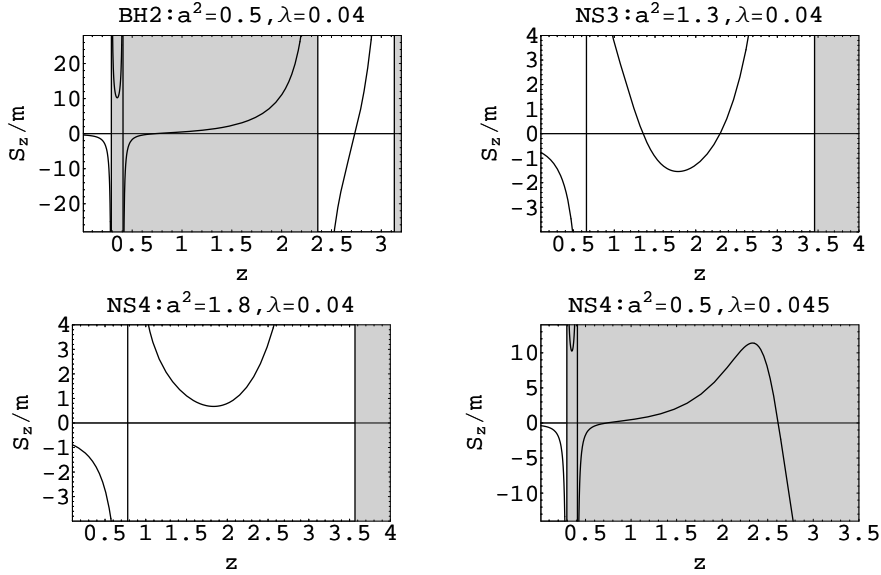}\end{center}
\caption{Behaviour of the function $S_{z}(z;a,\lambda)$. The function determines the magnitude and  orientation of the component of the spin vector directed along the axis of the spinning particle in equilibrium at the radius $z$, for given values of the parameters $a^2$ and $\lambda$ ($S_x=0$, $S_y=0$). The function vanishes at the static radii and diverges at the horizons, denoted by the boundary lines between the gray and white regions, and at $z^2=a^2/3$, denoted by the solid vertical line. We give here four examples of the behaviour of the function. Note that the function $S_{z}(z;a,\lambda)$ is not relevant in the dynamic regions (gray).}
\label{Fig:4}
\end{figure}

Thus, in the stationary regions, along the symmetry axis of the KdS spacetimes, the equilibrium of spinning test particles requires $S_z=const$ because of the relation (\ref{39}), and the values of $S_z$ are determined by the equation (\ref{44}). The equations (\ref{40}), (\ref{40a}), and (\ref{42})-(\ref{44}) can be satisfied under the following conditions. In the case of 
\begin{itemize}  
\item[$\bullet$] $S_x=0$, $S_y=0$, the conditions (\ref{42}) and (\ref{42a}) are automatically satisfied. The equilibrium is possible at all the radii satisfying the conditions (\ref{36a}) with spin given by the function $S_{z}(z;a,\lambda)$ determined by the relation (\ref{44}). For $S_z=0$, we obtain the equilibrium at the static radii given by solutions of the equation (\ref{45}).
Note that the function $S_{z}(z;a,\lambda)$ diverges at $z^2=a^2/3$, where there is no equilibrium possible. 
We give the function $S_{z}(z;a,\lambda)$ for a few specifically chosen values of the spacetime parameters in Figure \ref{Fig:4}. 
\item[$\bullet$] $S_x\neq0$, $S_y\neq0$, the conditions (\ref{42}) and (\ref{42a}) are satisfied only for $a^2=3z^2$. But there is no finite value for the spin $S_z$ given by the equation (\ref{44}), because the function $S_{z}(z;a,\lambda)$ diverges here, and the equilibrium does not occur in this case.
\end{itemize}
\subsubsection{Kerr and Schwarzschild-de Sitter spacetimes}
On the symmetry axis of the Kerr spacetimes, the equilibrium of spinning test particles is again  possible only in the stationary regions, i.e, in the regions where $r>2$. The limit case ($\lambda=0$) of the equilibrium conditions (\ref{39})-(\ref{40a}) and (\ref{42})-(\ref{44}) implies that the equilibrium requires $S_z=const$, as well as in the KdS spacetimes. In the case of
\begin{itemize}  
\item[$\bullet$] $S_x=0$, $S_y=0$, the conditions (\ref{42}) and (\ref{42a}) are automatically satisfied. Due to the condition (\ref{44}), the equilibrium is possible at all the radii satisfying the conditions (\ref{36a}), i.e., at $z>2$ with spin $S_{z}$ given by the relation 
\begin{eqnarray}
S_{z}=m\frac{(a^2+z^2)^2(z^2-a^2)}{2a(a^2-3z^2)\Delta_z}.
\end{eqnarray}
For $S_{z}=0$ we obtain the equilibrium at the radii $z^2=a^2$.  
Note that this equation is not satisfied for any $S_z$ in the case of $a^2=3z^2$ as well as in the KdS case.
\item[$\bullet$] $S_x\neq0$, $S_y\neq0$, the conditions (\ref{42}) and (\ref{42a}) are satisfied only for $a^2=3z^2$ and there is no solution for the finite values of the spin $S_z$, thus there is no equilibrium in this case. 
\end{itemize}

On the symmetry axis of the SdS spacetimes, the limit case ($a=0$) of the equilibrium conditions (\ref{39})-(\ref{40a}) and (\ref{42})-(\ref{44}) naturally implies that the equilibrium is possible only at the static radius as well as in the equatorial plane. 
\section{Conclusions}
In the KdS spacetimes, the combined effect of the rotation of the source and the cosmic repulsion enriches significantly the properties of the test particle equilibrium not only for the spinning particles, but also for the non-spinning particles, in contrast to the limiting spherically symmetric SdS case, where the 
equilibrium is spin independent and possible at the static radius only \cite{Stu:1999:ACTPS2:} (see Tables  \ref{Tab:2} and \ref{Tab:3}).

In the case of the non-spinning particles (the equilibrium of which is given by the geodetical structure of the spacetimes), the equilibrium position is allowed in the equatorial plane at the   rotational parameter independent static radius $r_{sr}=\lambda^{-1/3}$, formally coinciding with the SdS formula \cite{Stu:1999:ACTPS2:}. The equatorial equilibrium is possible at $r_{sr}$ for any KdS black-hole or naked-singularity admitting existence of the static limit surfaces (see Figure~\ref{Fig:2}), but it is not possible in any Kerr spacetime. Notice, however, that the particles in equilibrium in the equatorial plane are rotating relative to the locally non-rotating frames \cite{Stu-Sla:2004:PHYSR4:}. 

On the symmetry axis, the static radii depend on both the spacetime parameters and are implicitly given by the equation (\ref{45}), reduced in the SdS limit to the known result $r_{sr}=\lambda^{-1/3}$. There is one static radius in all black-hole spacetimes, while in naked-singularity spacetimes, even two static radii can appear. In the Kerr spacetimes, the equilibrium is possible at the radii $z^2=a^2$, i.e., it is allowed in the naked-singularity spacetimes, while it is forbidden in the black-hole spacetimes.

The equilibrium of the spinning particles is spin-dependent in contrast to the SdS case. 
In the equatorial plane of the KdS spacetimes allowing the existence of static limit surfaces, the equilibrium is possible at the static radius, if $S_\varphi=const\neq0$ and $S_r=S_\theta=0$ ($\varphi$-directed spin). If $S_\theta=const\neq0$ and $S_\varphi=S_r=0$ (equatorial plane orthogonal spin), the equilibrium is possible at the radii outside the ergosphere with the spin component $S_\theta$ determined by the equation (\ref{32}). 
In the Kerr spacetimes, the equilibrium is possible outside the ergosphere, but only in the case of the equatorial plane orthogonal spin given by the relation (\ref{35a}) .
\begin{table}
\caption{\label{Tab:2} Equilibrium positions of spinning test particles in the equatorial plane of the KdS, Kerr, and SdS spacetimes in dependence on the particle spin.}
\begin{indented}
\item\begin{tabular}{@{}c|c|c|c|c|c}
\br
$\bf{S_r}$&$\bf{S_\varphi}$&$\bf{S_\theta}$&\bf{SdS}&\bf{Kerr}&\bf{KdS}\\
\hline
=0&=0&=0&$r_{sr}$&-----&$r_{sr}$\\
&&$\neq0$&$r_{sr}$&$r=r(S_\theta,a)$&$r=r(S_\theta,a,\lambda)$\\
\hline
&$\neq0$&$=0$&$r_{sr}$&-----&$r_{sr}$\\
&&$\neq0$&$r_{sr}$&-----&-----\\
\hline
$\neq0$&arbitr.&arbitr.&$r_{sr}$&-----&-----\\
\br
\end{tabular}
\end{indented}
\end{table}

On the axis of symmetry, the equilibrium is allowed only for particles with $S_x=0$ and $S_y=0$ (the spin oriented along the axis) at any position in the stationary regions of the black-hole or naked-singularity spacetimes. At a given radius, the relevant component of the spin $S_z$ is determined by the equation (\ref{44}). Depending on the position, the spin can be oriented inward or outward, which happens, due to the cosmic repulsion, even in the black-hole spacetimes (see Figure \ref{Fig:4}). On the other hand, in the Kerr black-hole spacetimes, the spin is inward oriented everywhere since $z^2>a^2$, while in the Kerr naked-singularity spacetimes, it is inward oriented at $z^2>a^2$, outward oriented at $a^2/3<z^2<a^2$, and inward oriented at $z^2<a^2/3$. At the static radius on the symmetry axis of the KdS spacetimes, the spin of particles in equilibrium must be zero.
\begin{table}
\caption{\label{Tab:3} Equilibrium positions of spinning test particles on the symmetry axis of the  KdS, Kerr, and SdS spacetimes in dependence on the particle spin.}
\begin{indented}
\item\begin{tabular}{@{}c|c|c|c|c}
\br
$\bf{S_x}$ and $\bf{S_y}$&$\bf{S_z}$&\bf{SdS}&\bf{Kerr}&\bf{KdS}\\
\hline
=0&=0&$r_{sr}$&-----&$r_{sr}$\\
&$\neq0$&$r_{sr}$&$r=r(S_z,a)$&$r=r(S_z,a,\lambda)$\\
\hline
$\neq0$&arbitr&$r_{sr}$&-----&-----\\
\br
\end{tabular}
\end{indented}
\end{table}

\ack
This work was supported by the Czech grant MSM 4781305903.

\end{document}